\documentclass[preprints,article,accept,moreauthors,pdftex]{Definitions/mdpi}
\usepackage{graphicx}
\usepackage{caption}
\usepackage{subcaption}

\firstpage{1} 
\pubvolume{xx}
\issuenum{1}
\articlenumber{5}
\pubyear{2019}
\copyrightyear{2019}
\history{}

\Title{Study of angular correlations in Monte Carlo simulations}

\Author{Balázs Endre Szigeti $^{1,2}$*,Mónika Varga-Kofarago $^{2}$*}

\AuthorNames{Firstname Lastname, Firstname Lastname and Firstname Lastname}

\address{%
$^{1}$ \quad Eötvös Loránd University, Department of Atomic Physics, Budapest, Hungary\\
$^{2}$ \quad HAS Wigner Research Centre for Physics, Institute for Particle and Nuclear Physics, Budapest, Hungary}

\corres{Correspondence: szigeti.balazs@wigner.mta.hu, varga-kofarago.monika@wigner.mta.hu}

\abstract{In two-particle angular correlation measurements the distribution of charged hadron pairs are evaluated as a function of pseudorapidity ($\Delta \eta$) and azimuthal ($\Delta \varphi$) differences. In these correlations, jets manifest themselves as a near-side peak around $\Delta \eta = 0,$  $\Delta \varphi = 0$.  These correlations can be used to extract transverse momentum ($p_T$) and centrality dependence of the shape of the near-side peak in Pb-Pb collision. The shape of the near-side peak is quantified by the variances of the distribution. The variances are evaluated from a fit combining the peak and the background. In this contribution, identified and unidentified angular correlations are shown from Pb-Pb collisions  at $\sqrt{s_{NN}} = 2.76$ TeV from Monte Carlo simulations (AMPT, PYTHIA 8.235/Angantyr). Results show that transport models in AMPT give better results than PYTHIA 8.235/Angantyr when comparing to the experimental results of the ALICE collaboration.}


\begin{document}


\section{Introduction}

In heavy-ion collisions where the typical momentum transfer is large $(Q^2 >> \Lambda_{QCD})$ partons with high transverse momentum ($p_T$) are produced. Since the partons carry color charge, they can not exist freely at low energies \cite{Gross}. For this reason they eventually hadronize into a shower of correlated hadrons called jets. These jets are produced in the early stage of the collisions and they propagate through the Quark-Gluon Plasma. During the propagation they interact with the hot and dense medium. Jets lose momentum, due to a chain of elementary processes, for instance induced gluon radiation and elastic scattering. These processes  are jointly referred to as jet quenching \cite{Levai}. Experimental studies of these high $p_T$ yields can be used to study the generated medium. Beside  the experimental studies Monte Carlo based simulations are commonly used to model heavy-ion collision systems. \\

In the ALICE experiment \cite{ALICE}, unidentified hadron-hadron angular correlations are used to analyze $\mathrm{PbPb}$ and $\mathrm{pp}$ collision at $\sqrt{s_{NN}}= 2.76$ TeV in the $p_T$ region of $1$ GeV/\emph{c} $< p_T < 8$ GeV/\emph{c}. The results show that in high centrality and low $p_T$ cases the jet-peak becomes wider and asymmetrical in $\Delta \eta, \Delta \varphi$. Furthermore, a depletion appears around $\Delta \varphi = 0$, $\Delta \eta = 0$. Both the jet broadening and the depletion are seen in AMPT simulations carried out by the ALICE Collaboration ~\cite{ANL1,ANL2}.\\

In this contribution, identified and unidentified angular correlations are shown from Pb-Pb collisions from Monte Carlo simulations to study which type of particles show similar properties as measured by the ALICE experiment. Moreover, we used Monte Carlo simulations with different physical assumption to study which physical processes are responsible for the observed phenomena.

\section{Analysis Method}

\begin{figure}[ht]
	\centering
	\begin{subfigure}{0.45\textwidth}
		\includegraphics[width=\textwidth]{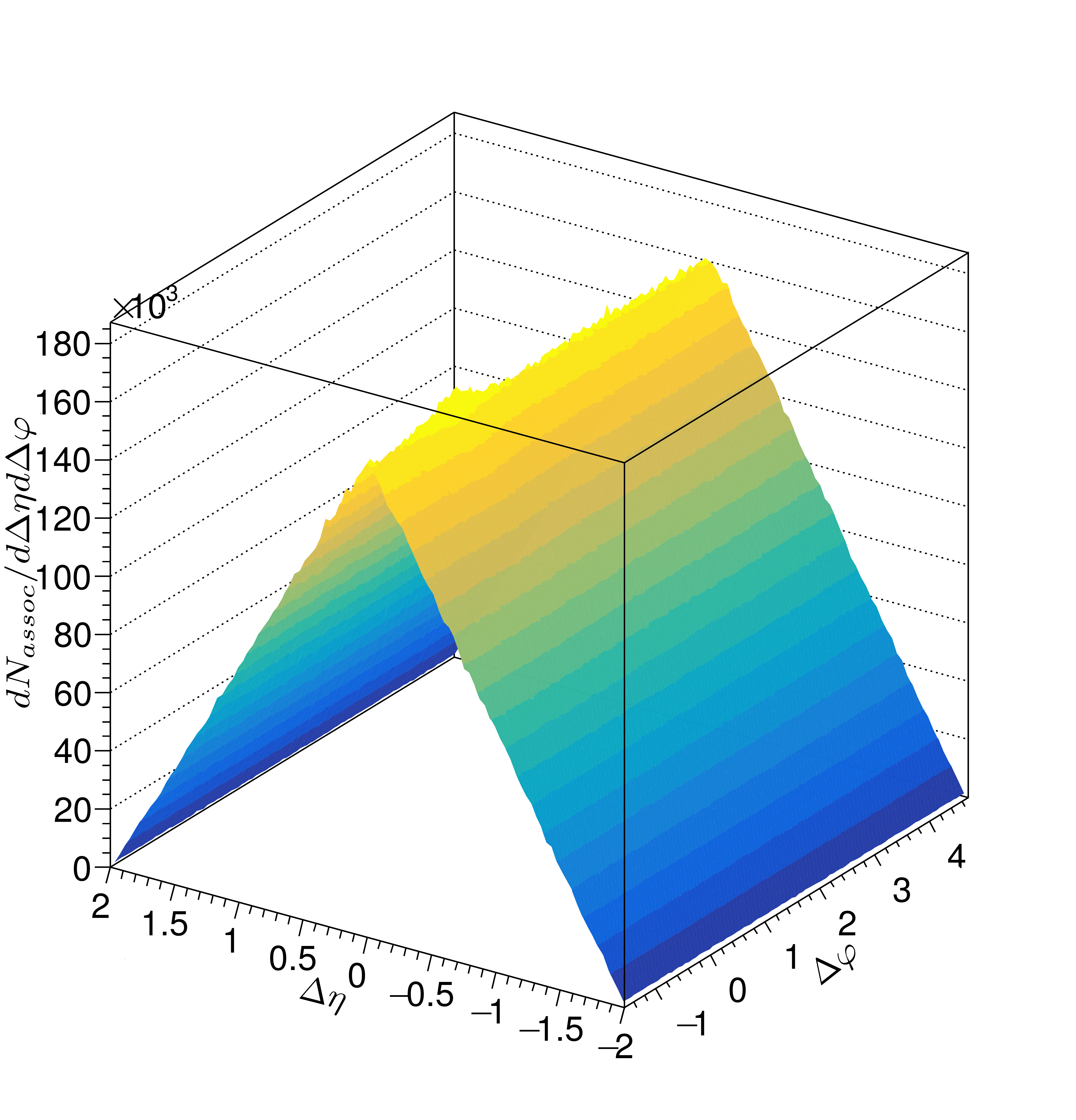}
	\end{subfigure}
	\vspace{1em}
	\begin{subfigure}{0.45\textwidth}
		\includegraphics[width=\textwidth]{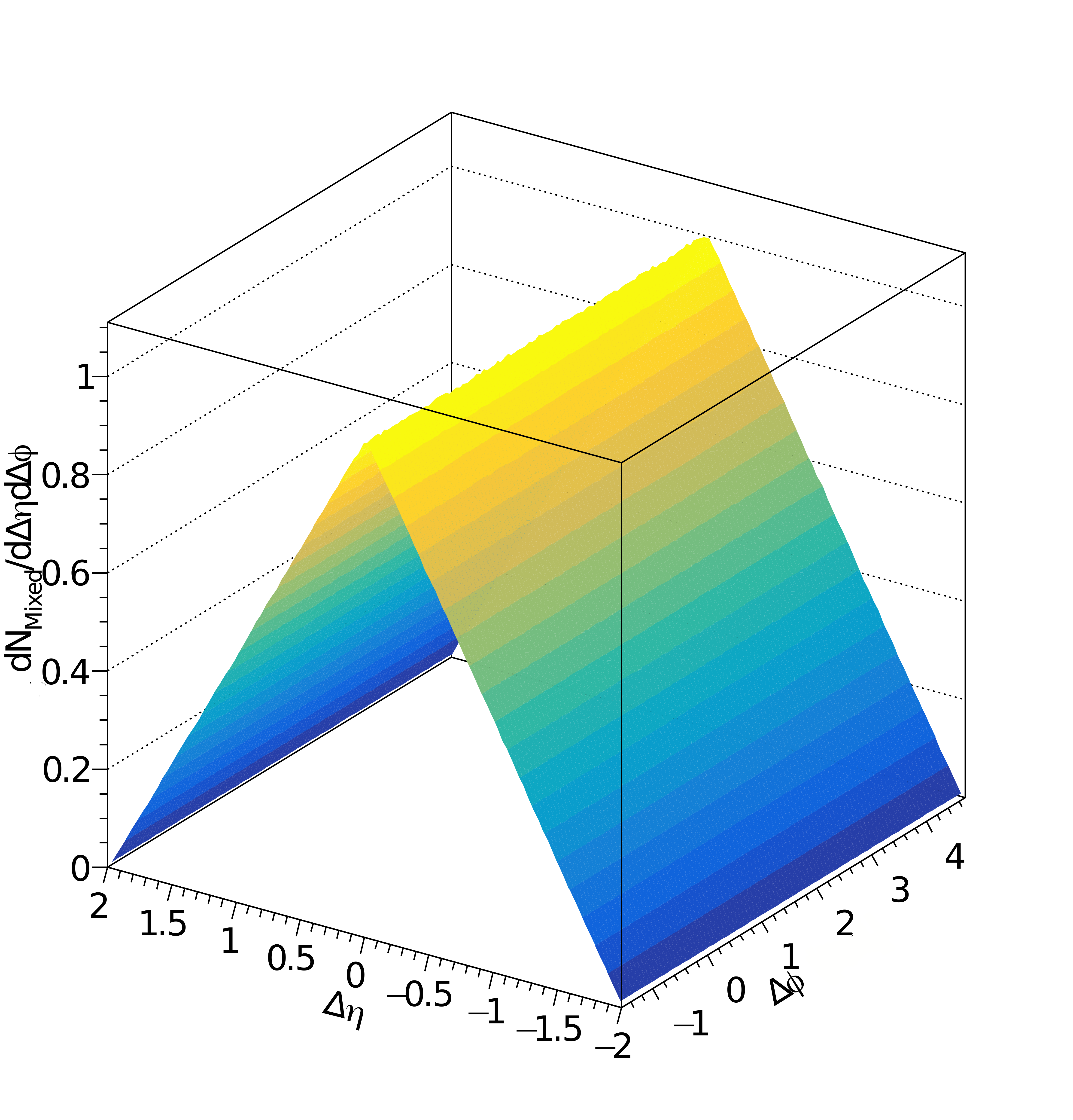}
	\end{subfigure}
	\caption{The $S(\Delta \varphi, \Delta \eta)$ (left) and the $M(\Delta \varphi, \Delta \eta)$ (right) distributions in a typical $\mathrm{Pb-Pb}$ sample in AMPT at $\sqrt{s_{NN}}= 2.76$ TeV, where transverse momentum are \break $1$ GeV/\emph{c} $< \mathrm{p}_{\mathrm{T,assoc}},\mathrm{p}_{\mathrm{T,trigg}} < 2$ GeV/\emph{c} and centrality is between 10-20\%.}
	\label{fig::pt2}
\end{figure}

\begin{figure}[ht]
	\centering
	\begin{subfigure}{0.45\textwidth}
		\includegraphics[width=\textwidth]{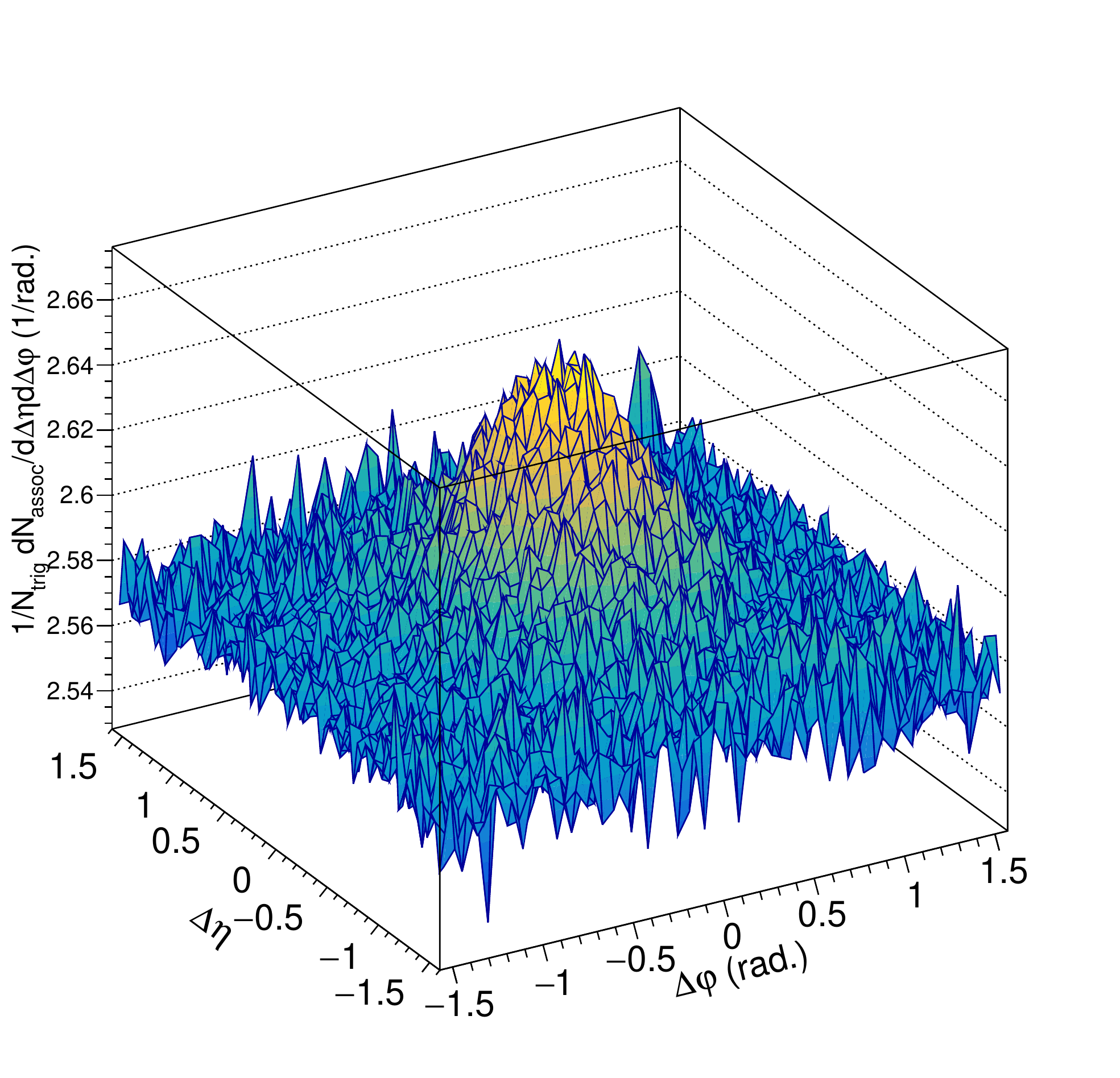}
	\end{subfigure}
	\vspace{1em}
	\begin{subfigure}{0.45\textwidth}
		\includegraphics[width=\textwidth]{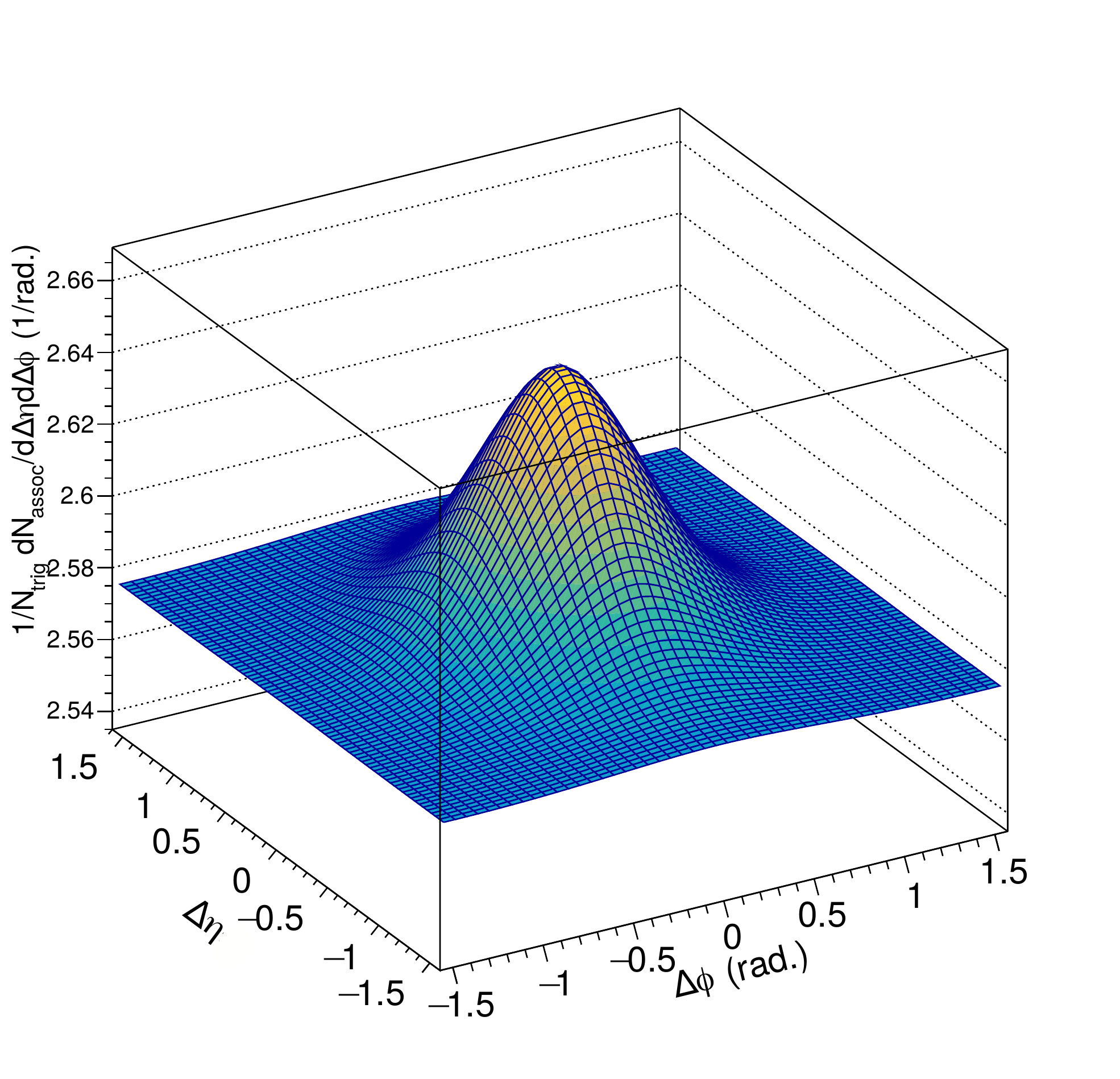}
	\end{subfigure}
	\caption{The near-side jet peak (left) and the generalized Gaussian fit (right) in a typical $\mathrm{Pb-Pb}$ sample in AMPT at $\sqrt{s_{NN}}= 2.76$ TeV, where transverse momentum are \break $1$ GeV/\emph{c} $< \mathrm{p}_{\mathrm{T,assoc}},\mathrm{p}_{\mathrm{T,trigg}} < 2$ GeV/\emph{c} and centrality is between 10-20\%}
	\label{fig::pt3}
\end{figure}

In this work, (identified)hadron-hadron angular correlation measurements are used to compare Monte Carlo simulations with different physical processes. The correlation between particles is measured as a function of pseudorapidity $(\Delta \eta)$ and azimuthal $(\Delta \varphi)$ differences of trigger and associated particles. In every event we choose a particle (trigger), then the $(\Delta \eta)$ and $(\Delta \varphi)$ differences to other particles (associated) are evaluated. This process is then repeated for every particle as trigger from the particular event. Trigger and associated particles are chosen from trigger and associated transverse momentum intervals ($\mathrm{p}_{\mathrm{T,trig}}$, $\mathrm{p}_{\mathrm{T,assoc}}$), respectively. The intervals can be disjoint or identical, in the second case only those pairs will be considered, where $p_{\mathrm{T,trig}} > p_{\mathrm{T,assoc}}$ to avoid double counting. At the experiment the detector acceptance is limited in $|\eta|$, due to this reason $\Delta \eta$ is restricted to $|\Delta \eta|<2$. The associated yield per-trigger can be expressed in terms of the ratio:

\begin{equation}
    \dfrac{1}{N_{trig}}\dfrac{d^2N_{assoc}}{d\Delta \varphi d\Delta \eta} = \dfrac{S(\Delta \varphi,\Delta \eta)}{\alpha M(\Delta \varphi,\Delta \eta)},
\end{equation}
 
where $S(\Delta \varphi, \Delta \eta)$ is the signal distribution, where the trigger and associated particles are chosen from the same event normalized by the total number of triggers (Fig.~\ref{fig::pt2}). This is in contrast with $M (\Delta \varphi, \Delta \eta)$ where particles are from separate events scaled by $\alpha$, the value of $M(\Delta \varphi,\Delta \eta)$ in (0,0). This ratio corrects for pair inefficiencies and detector acceptance effects \cite{ANL1}. In addition to this, the effects from long-lived neutral-particle decays ($K^0_s$ and $\Lambda^0$) and $\gamma$-conversion are removed by cutting on the invariant mass ($m_{inv}$) of particle pairs. Although, in simulation the cut can be applied directly on the decay products, as particle information is stored during the hadron shower, we would like to maintain consistency with the experimental analysis for better comparison (Fig. \ref{fig::pt3}).\\

In order to characterize the obtained associated yield per-trigger a combined fit is performed (Fig.~\ref{fig::pt3}). The shape of the near-side jet peak is parameterized by the variances  $\sigma_{\Delta \varphi}$ and $\sigma_{\Delta \eta}$  from the fit by a two-dimensional generalized Gaussian function (Eq.~\ref{eq:Gauss}). 

\begin{equation} \label{eq:Gauss}
	G_{\gamma,\omega}(\Delta \varphi,\Delta \eta) = N\dfrac{\gamma_{\Delta \varphi}\gamma_{\Delta \eta}}{4\omega_{\Delta \varphi}\omega_{\Delta \eta} \Gamma(1/\gamma_{\Delta \varphi})\Gamma(1/\gamma_{\Delta \eta})} exp \bigg[ - \bigg( \dfrac{\mid \Delta \varphi \mid}{\omega_{\Delta \varphi}} \bigg)^{\gamma_{\Delta \varphi}}- \bigg( \dfrac{\mid \Delta \eta \mid}{\omega_{\Delta \eta}} \bigg)^{\gamma_{\Delta \eta}} \bigg]
\end{equation}

The generalized Gaussian distribution equals to a standard Gaussian if $\gamma = 2$, while it is an exponential distribution if  $\gamma = 1$. The $\sigma_{\Delta \varphi, \Delta \eta}$ variances are evaluated form the $\omega$ and $\gamma$ parameters in the following way.

\begin{equation}  \label{eq:Sigma}
	\sigma_{\Delta \varphi; \Delta \eta} = \sqrt{\dfrac{\omega_{\Delta \varphi;\Delta \eta}^2 \Gamma(3/\gamma_{\Delta \varphi;\Delta \eta})}{\Gamma(1/\gamma_{\Delta \varphi;\Delta \eta})}}
\end{equation}

In Pb-Pb collisions, long-range correlations come from collective effects, where one of the essential elements of this is azimuthal flow. The background is characterized by a $C_1$ constant and the $V_{n\Delta}$ parameters which are Fourier components describing the azimuthal flow.  

\begin{equation} \label{eq:Fourier}
    F(\Delta \varphi, \Delta \eta) = C_1 + \sum_{n=2}^4 2V_{n\Delta}\mathrm{cos}(n\Delta \varphi)
\end{equation}

Due to the available simulation statistics the fit was performed separately on the different projections, instead of fitting in two dimensions. Hence, we lose some information about the shape of the jet-peak, but the accuracy of the fit increases.

\section{Results}

In this work, PYTHIA 8.235/Angantyr \cite{PYTHIA,Angantyr} and AMPT \cite{AMPT} are used to simulate Pb-Pb collisions at $\sqrt{s_{NN}} = 2.76$ TeV. PYTHIA was developed to simulate $\mathrm{pp}$, $\mathrm{p\overline{p}}$ and $\mathrm{e^+e^-}$ collision. Recently a new heavy-ion simulation model has been built on PYTHIA 8.235, called Angantyr. Angantyr is based on the Fritiof model \cite{Fritiof}, and also contains MPI and CR models from PYTHIA \cite{PYTHIA}. AMPT is based on HIJING and the PYTHIA/JETSET models, and it is directly developed for simulating heavy-ion collisions. In AMPT, the collective effects are described by transport models. In this model, two main settings can be switch on/off: string melting and hadronic rescattering.\\

\begin{figure}[ht]
	\centering
	\begin{subfigure}{0.48\textwidth}
		\includegraphics[width=\textwidth]{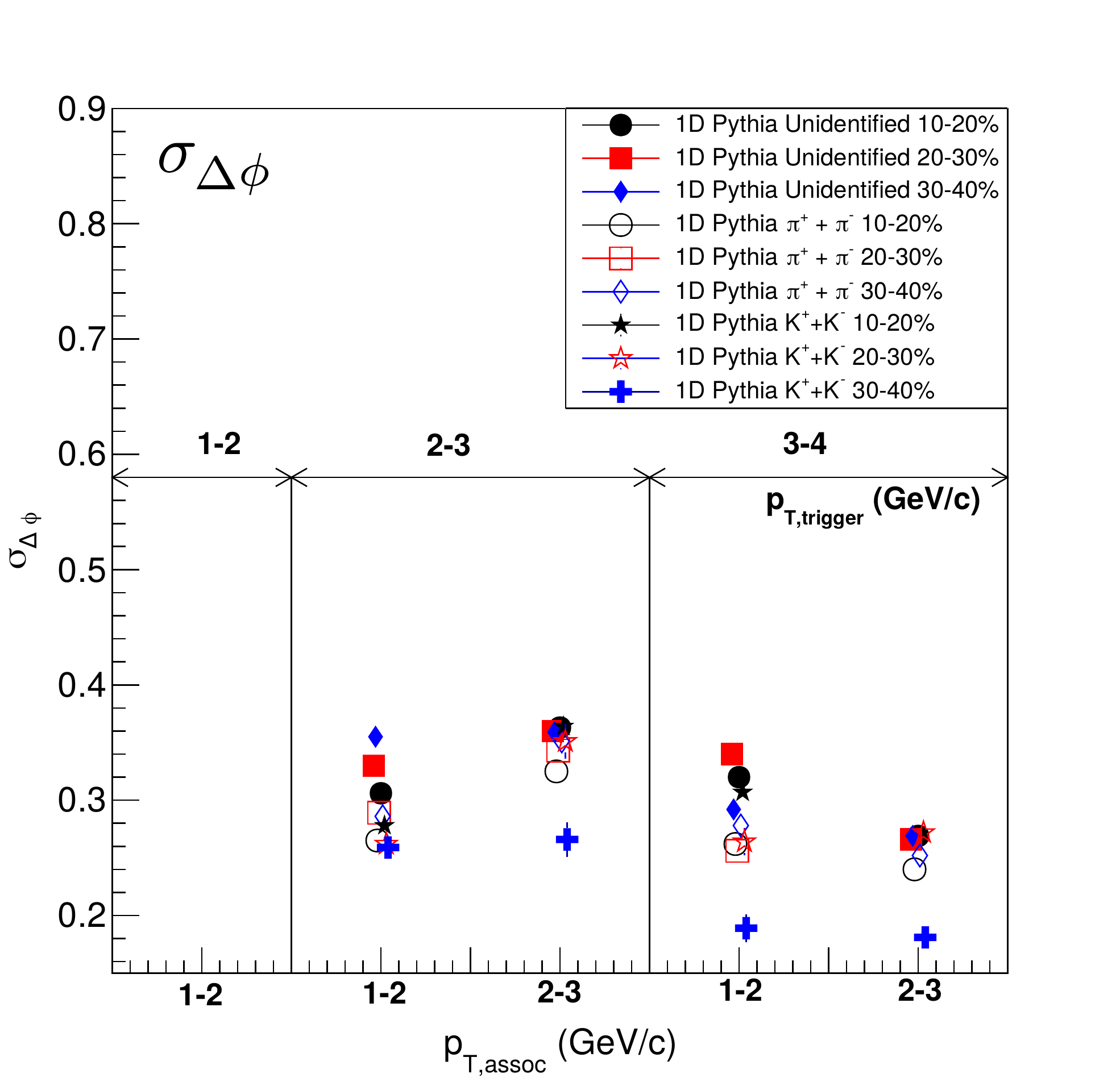}
	\end{subfigure}
	\vspace{1em}
	\begin{subfigure}{0.48\textwidth}
		\includegraphics[width=\textwidth]{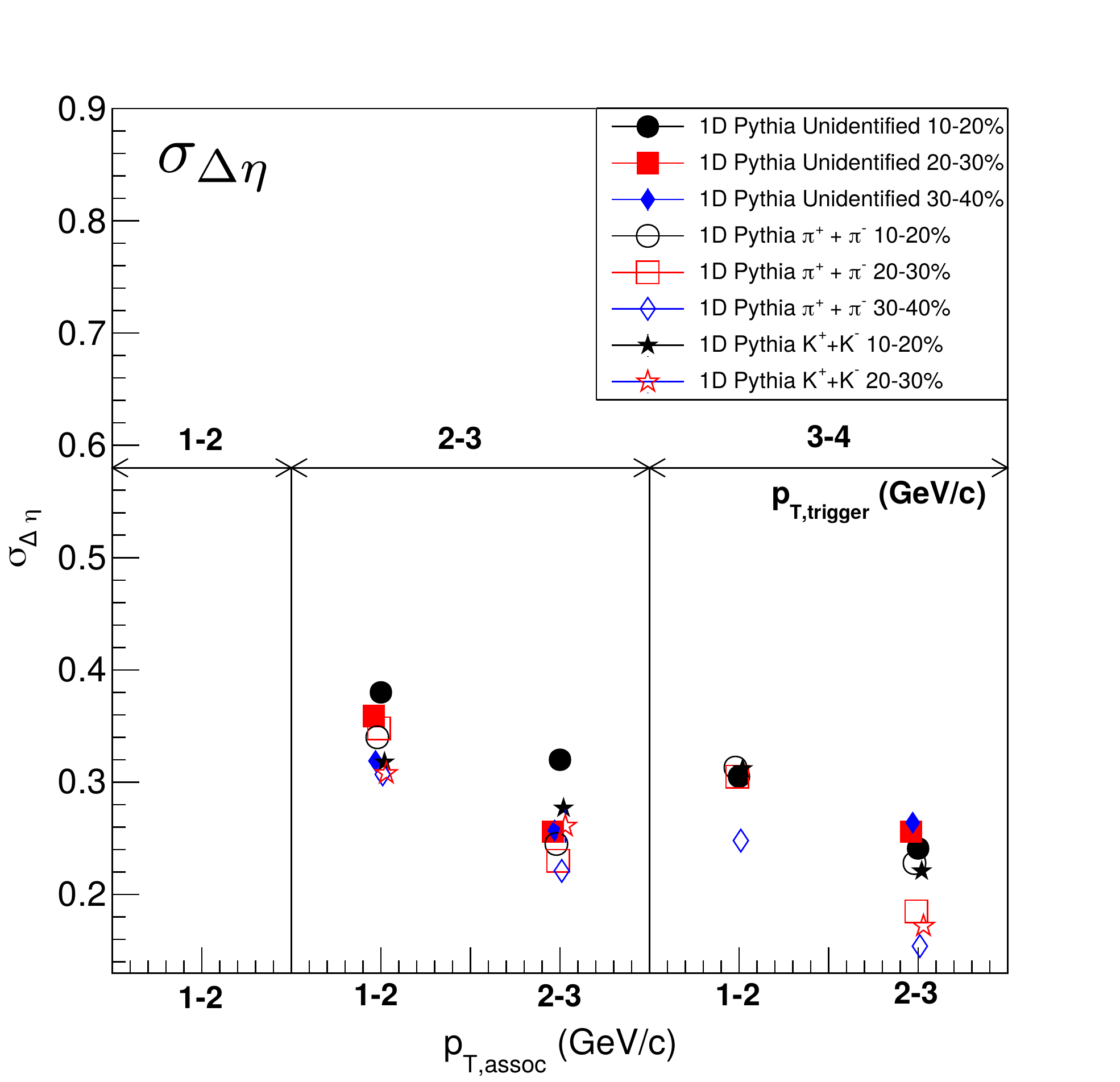}
	\end{subfigure}
	\caption{Comparison of the $\sigma_{\Delta \eta}$ and $\sigma_{\Delta \varphi}$ variances from PYTHIA 8.235/Angantyr}
	\label{fig::PYTHIA01}
\end{figure}

Unidentified and identified (where trigger particles are kaons or pions) angular correlation are compared at three different centrality bins from PYTHIA 8.235/Angantyr. In the \break $1$ GeV/\emph{c} $< p_T < 2$ GeV/\emph{c} transverse momentum bin the near-side jet peak was not significant enough for the fit. The obtained variances are depicted in Fig.~\ref{fig::PYTHIA01}. The figure shows that there is a particle species dependence in PYTHIA. 

\begin{figure}[ht]
	\centering
	\begin{subfigure}{0.48\textwidth}
		\includegraphics[width=\textwidth]{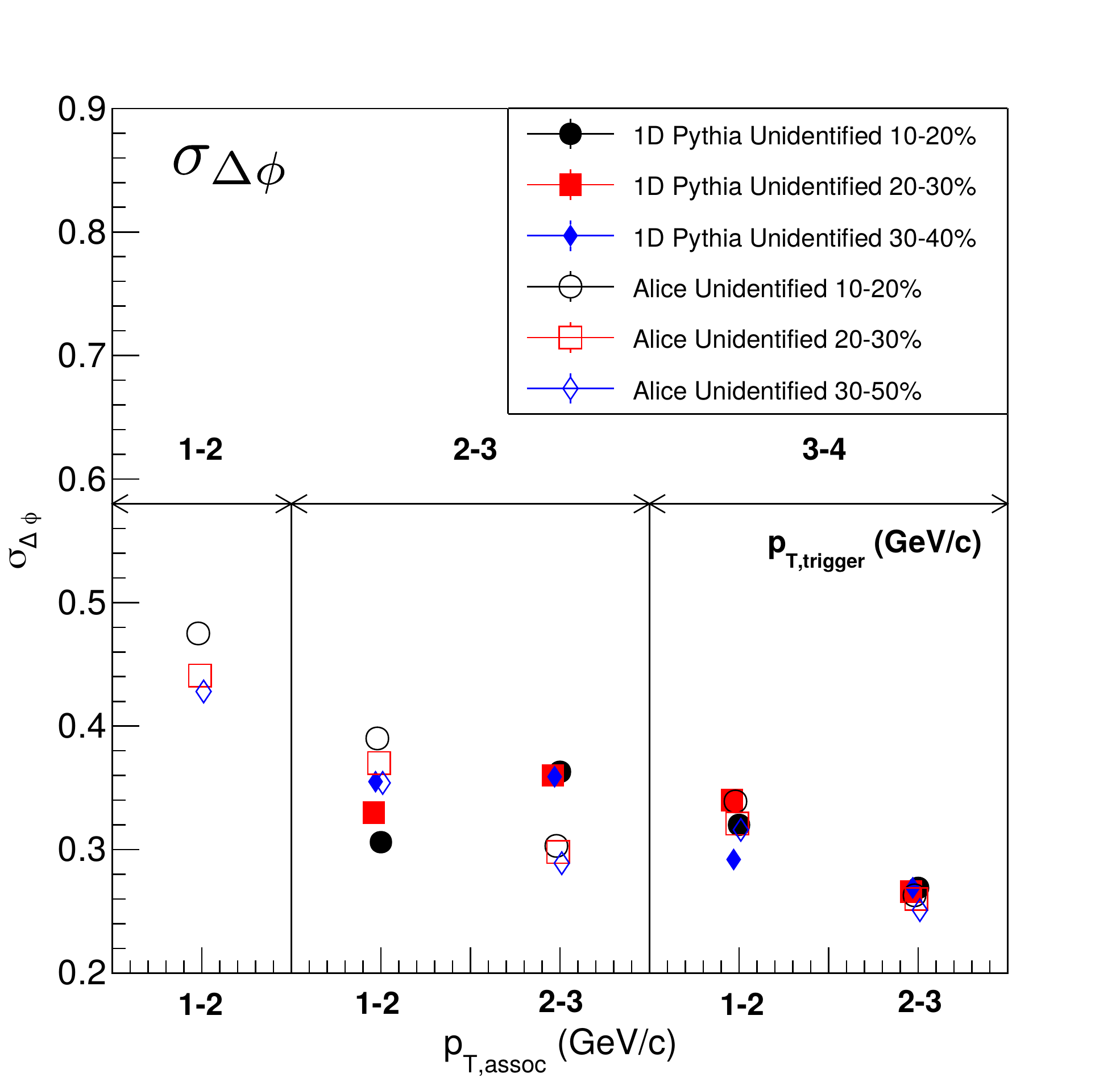}
	\end{subfigure}
	\vspace{1em}
	\begin{subfigure}{0.48\textwidth}
		\includegraphics[width=\textwidth]{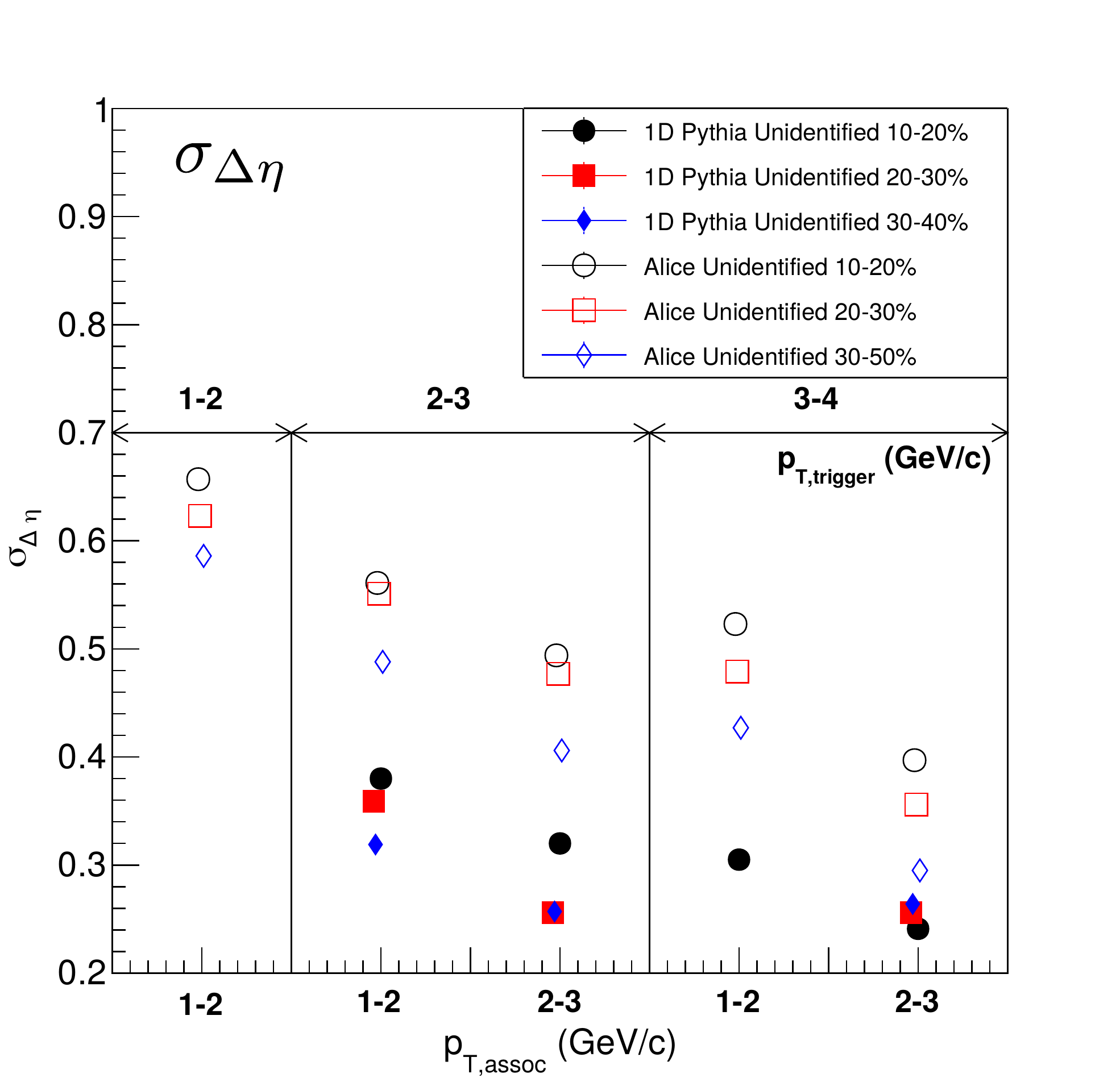}
	\end{subfigure}
	\caption{Comprasion of the $\sigma_{\Delta \eta}$ and $\sigma_{\Delta \varphi}$ variances from PYTHIA and ALICE measurements \cite{ANL1}.}
	\label{fig::PYTHIA02}
\end{figure}

The PYTHIA 8.235/Angantyr results are compared with the unidentified hadron-hadron correlation from the ALICE experiment. These can be seen in Fig.~\ref{fig::PYTHIA02}. It is clearly visible that PYTHIA gives a better description in the $\Delta \varphi$ direction than in the $\Delta \eta$ direction. However, it does not describe the data well in any of the directions. 

\begin{figure}[ht]
	\centering
	\begin{subfigure}{0.48\textwidth}
		\includegraphics[width=\textwidth]{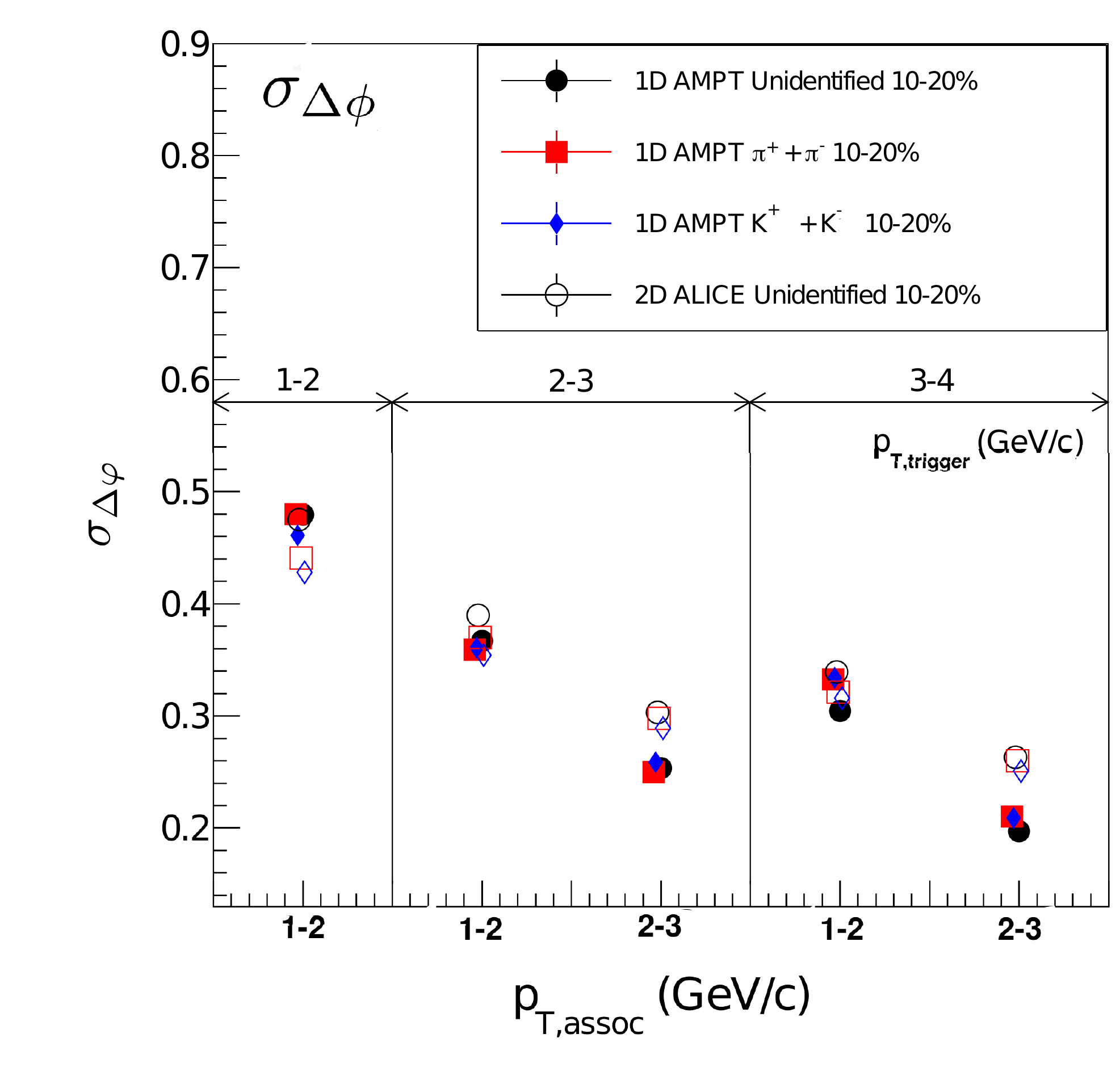}
	\end{subfigure}
	\vspace{1em}
	\begin{subfigure}{0.48\textwidth}
		\includegraphics[width=\textwidth]{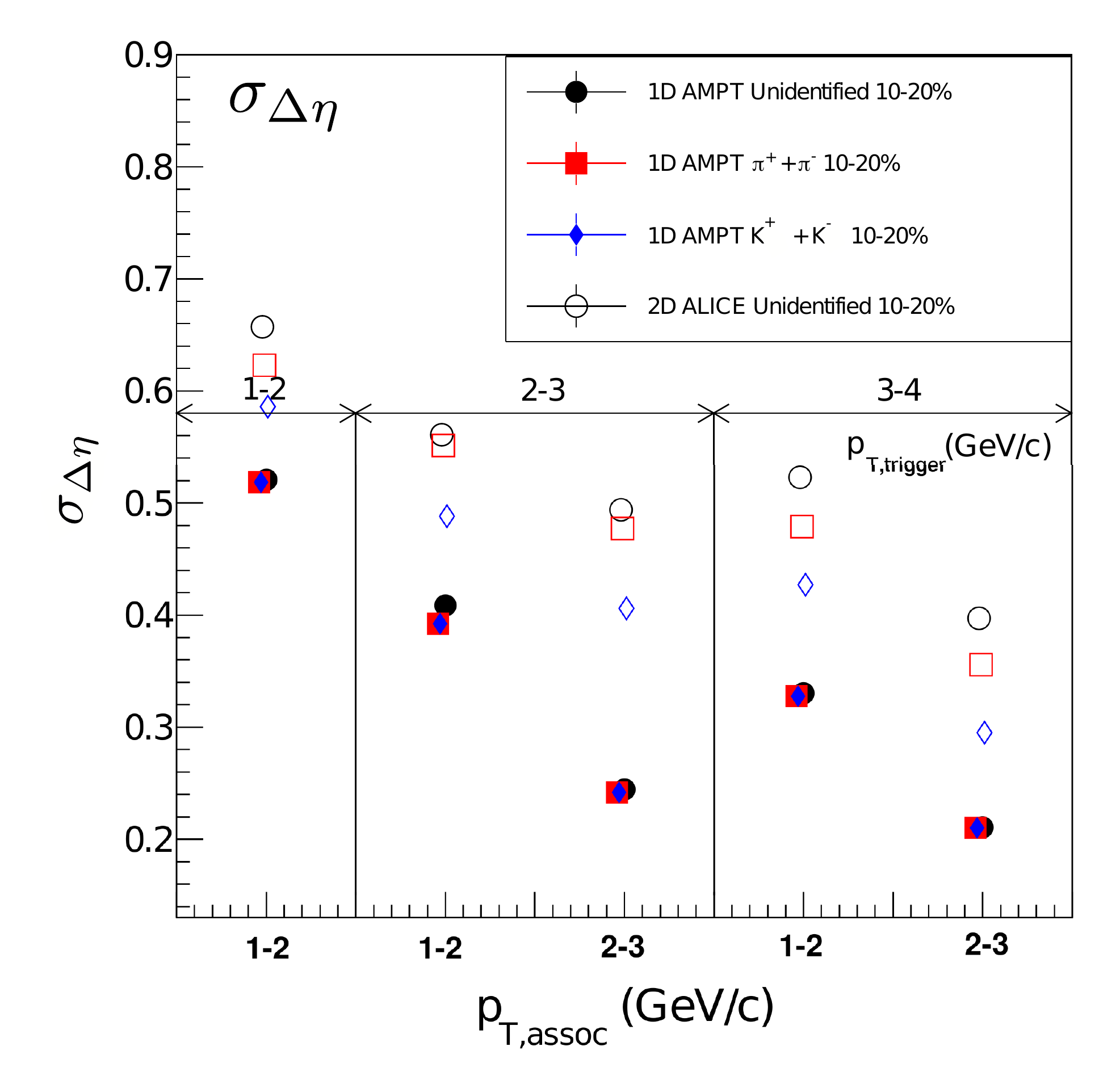}
	\end{subfigure}
	\caption{Comparison of the $\sigma_{\Delta \eta}$ and $\sigma_{\Delta \varphi}$ variances from AMPT to the ALICE measurements \cite{ANL1}.}
	\label{fig::AMPT01}
\end{figure}

In the case of AMPT, only the most central bin has been simulated due to the reason that the AMPT simulation time is significantly longer than in the case of PYTHIA 8.235/Angantyr. The obtained $\sigma_{\Delta \varphi}$ and $\sigma_{\Delta \eta}$ variances from AMPT\footnote{String melting: on, Hadronic rescattering: off, Lund String fragmentation parameters: 0.30 GeV/$c^2$, 0.15 GeV/$c^2$, The $p_t$ cutoff for minijets: 2.0 GeV/\emph{c}, Quenching flag: off, Shadowing flag: on, Parton screening mass: 2.265 fm$^{-1}$}. are compared to the ALICE experimental results and to the AMPT results simulated by ALICE collaboration (Fig.~\ref{fig::AMPT01}). The figure shows that AMPT gives better description of the widths in $\Delta \varphi$, than in $\Delta \eta$. Furtermore, AMPT gives better trends compared to the experimental results, than PYTHIA 8.235/Angantyr. It is visible that the parameter set used by ALICE to simulate Pb-Pb events overestimates the widths, contrary to our parameter set which underestimates the widths in the $\Delta \eta$ direction. Consequently, there is probably an ideal parameter set where the widths from AMPT fit well in both directions. The difference between the widths from unidentified and identified measurements is not as significant as in PYTHIA 8.235/Angantyr.  

\section{Conclusion}

To summarize, we used unidentified and identified hadron-hadron angular correlations to analyze Monte Carlo simulations. We used a combined fit to characterize the obtained near-side jet peaks. We have seen a hint of particle species dependence in PYTHIA 8.235/Angantyr. As PYTHIA does not describe well the jet-peak widths measured by the ALICE collaboration, we can conclude that MPI and CR models can not describe properly the collective processes seen in Pb-Pb collisions. Furthermore, we found that AMPT gives better results, than PYTHIA 8.235/Angantyr when compared to the experimental results. However, we note that it is difficult to find the ideal parameter set in AMPT simulation, due to the reason that the parameter set is large.  

\funding{This research has been supported by the Hungarian NKFIH/OTKA K 120660 grant.}

\conflictsofinterest{The authors declare no conflict of interest. The funders had no role in the design of the study; in the collection, analyses, or interpretation of data; in the writing of the manuscript, or in the decision to publish the results.}
\vspace{6pt} 
\noindent
\section{Abbriviations}
\begin{tabular}{@{}ll}
ALICE & A Large Ion Collider Experiment\\
QCD & Quantum chromodynamics\\
AMPT & A Multi-Phase Transport Model\\
MPI & Multi-Parton Interaction\\
CR & Color Reconnection \\
\end{tabular}


\reftitle{References}





\end{document}